\documentstyle[amsmath,graphicx]{mn}
\newif\ifAMStwofonts
\ifoldfss
  \ifCUPmtlplainloaded \else
    \NewTextAlphabet{textbfit} {cmbxti10} {}
    \NewTextAlphabet{textbfss} {cmssbx10} {}
    \NewMathAlphabet{mathbfit} {cmbxti10} {} % for math mode
    \NewMathAlphabet{mathbfss} {cmssbx10} {} %  "   "    "
  \fi
  \ifAMStwofonts
    \ifCUPmtlplainloaded \else
      \NewSymbolFont{upmath} {eurm10}
      \NewSymbolFont{AMSa} {msam10}
      \NewMathSymbol{\upi}     {0}{upmath}{19}
      \NewMathSymbol{\umu}     {0}{upmath}{16}
      \NewMathSymbol{\upartial}{0}{upmath}{40}
      \NewMathSymbol{\leqslant}{3}{AMSa}{36}
      \NewMathSymbol{\geqslant}{3}{AMSa}{3E}

      \let\leq=\leqslant \let\le=\leqslant
      \let\geq=\geqslant 
    \fi
  \fi
\fi % End of OFSS

\ifnfssone
  \newmathalphabet{\mathit}
  \addtoversion{normal}{\mathit}{cmr}{m}{it}
  \addtoversion{bold}{\mathit}{cmr}{bx}{it}
  \newmathalphabet{\mathbfit} % math mode version of \textbfit{..}
  \addtoversion{normal}{\mathbfit}{cmr}{bx}{it}
  \addtoversion{bold}{\mathbfit}{cmr}{bx}{it}
  \newmathalphabet{\mathbfss} % math mode version of \textbfss{..}
  \addtoversion{normal}{\mathbfss}{cmss}{bx}{n}
  \addtoversion{bold}{\mathbfss}{cmss}{bx}{n}
  \ifAMStwofonts
    \ifCUPmtlplainloaded \else
      \UseAMStwoboldmath
      \makeatletter
      \new@mathgroup\upmath@group
      \define@mathgroup\mv@normal\upmath@group{eur}{m}{n}
      \define@mathgroup\mv@bold\upmath@group{eur}{b}{n}
      \edef\UPM{\hexnumber\upmath@group}
      \new@mathgroup\amsa@group
      \define@mathgroup\mv@normal\amsa@group{msa}{m}{n}
      \define@mathgroup\mv@bold\amsa@group{msa}{m}{n}
      \edef\AMSa{\hexnumber\amsa@group}
      \makeatother
      \mathchardef\upi="0\UPM19
      \mathchardef\umu="0\UPM16
      \mathchardef\upartial="0\UPM40
      \mathchardef\leqslant="3\AMSa36
      \mathchardef\geqslant="3\AMSa3E

      \let\leq=\leqslant \let\le=\leqslant
      \let\geq=\geqslant 
    \fi
  \fi
\fi % End of NFSS release 1

\ifnfsstwo
  \DeclareMathAlphabet{\mathbfit}{OT1}{cmr}{bx}{it}
  \SetMathAlphabet\mathbfit{bold}{OT1}{cmr}{bx}{it}
  \DeclareMathAlphabet{\mathbfss}{OT1}{cmss}{bx}{n}
  \SetMathAlphabet\mathbfss{bold}{OT1}{cmss}{bx}{n}
  \ifAMStwofonts
    \ifCUPmtlplainloaded \else
      \DeclareSymbolFont{UPM}{U}{eur}{m}{n}
      \SetSymbolFont{UPM}{bold}{U}{eur}{b}{n}
      \DeclareSymbolFont{AMSa}{U}{msa}{m}{n}
      \DeclareMathSymbol{\upi}{0}{UPM}{"19}
      \DeclareMathSymbol{\umu}{0}{UPM}{"16}
      \DeclareMathSymbol{\upartial}{0}{UPM}{"40}
      \DeclareMathSymbol{\leqslant}{3}{AMSa}{"36}
      \DeclareMathSymbol{\geqslant}{3}{AMSa}{"3E}

      \let\leq=\leqslant \let\le=\leqslant
      \let\geq=\geqslant 
    \fi
  \fi
\fi % End of NFSS release 2
\ifCUPmtlplainloaded \else
  \ifAMStwofonts \else % If no AMS fonts
    \def\upi{\pi}
    \def\umu{\mu}
    \def\upartial{\partial}
  \fi
\fi
\title{A relativistic jet from Cygnus X-1 in the low/hard X-ray state}
\author[A. M. Stirling et al.]
       {A. M. Stirling$^{1\&2}$,  R. E. Spencer$^2$, C. J. de la Force$^2$, M. A. Garrett$^3$,
\newauthor  R. P. Fender$^4$ and R. N. Ogley$^5$\\
$^1$CFA, University of Central Lancashire, Preston, PR1 2HE, UK.\\
$^2$University of Manchester, Jodrell Bank Observatory, Macclesfield, Cheshire SK11 9DL, UK.\\ 
$^3$Joint Institute for VLBI in Europe, Postbus 2, 7990 AA Dwingeloo, NL.\\
$^4$Astronomical Institute 'Anton Pannekoek' and Centre for High--Energy Astrophysics, University of Amsterdam, \\Kruislaan 403, 1098 SJ Amsterdam, The Netherlands.\\
$^5$Physics Department, Keele University, Staffordshire, ST5 5BG, UK.
}
\date{Accepted 0000.
      Received 1999;
      in original form 1999 June 24}
\pagerange{\pageref{firstpage}--\pageref{lastpage}}
\pubyear{1994}
\begin{document}

\maketitle

\begin{abstract}

We present the detection of a radio-emitting jet from the black-hole
candidate and X-ray binary source Cygnus X-1. Evidence of a bright
core with a slightly extended structure was found on milliarcsecond
resolution observations with the VLBA at 15.4 GHz. Later observations
with the VLBA (and including the phased up VLA) at 8.4 GHz show an
extended jet-like feature extending to $\sim$ 15 mas from a core
region, with an opening angle of $<$ 2$^{\circ}$. In addition, lower
resolution MERLIN observations at 5 GHz show that the source has $<$
10 per cent linear polarization. The source was in the low/hard X-ray
state during the observations, and the results confirm the existence
of persistent radio emission from an unresolved core and a variable
relativistic ($>$~0.6~c) jet during this state.

\end{abstract}

\section{Introduction}
Of the $\sim$250 currently known X-ray binaries, approximately 50 have
detectable radio emission and a dozen or so of these have been found
to have radio jets. Extreme examples of the jet sources are
GRS~1915+105 (Mirabel \& Rodriguez 1994; Fender et al. 1999) and
GRO~J1655--40 (Hjellming \& Rupen 1995; Tingay et al. 1995), both
exhibiting apparent superluminal motion. Such jet sources often show
strong radio flux variability (see Fender, Burnell \& Waltman 1997 for
a review). However, some sources show relatively little variation,
such as Cygnus~X-1 and LS 5039. The persistently radio-emitting
(20--40 mJy) LS 5039, which is at a similar distance, has been shown
to have a milliarcsecond-scale jet (Parades et al. 2000).

Cygnus X-1 (HDE 226868, V1357 Cygni) comprises a supergiant secondary
(spectral type O9.7 Iab) with mass between 20--33 M$_{\odot}$,
together with a compact primary which is a strong black-hole candidate
with mass between 7--16 M$_{\odot}$. The orbital period has been
detected in optical, UV, X-ray and radio wavelengths (Pooley, Fender
\& Brocksopp 1999; Brocksopp et al. 1999b). The orbital period is $\sim$5.6
days  orbital radius of
0.2 AU is derived; at an assumed distance of 2 kpc (Gierli\'{n}ski et
al. 1999) this corresponds to an angular scale of 0.1 milliarcsecond
(mas). An orbital ephemeris has been derived from optical radial
velocity data (LaSala et al. 1998) and compared with other wavelength
data (Brocksopp et al. 1999b).

The source has been fairly stable in the low/hard X-ray state, at
around 15 mJy at cm wavelengths and with an extremely flat radio
spectrum. The radio emission was first observed following a `turn on'
corresponding to a X-ray state change in 1972. The `turn on' was
probably the recovery from a high/soft state during which time the jet
had been quenched as Zhang et al. (1997) show that the radio emission
was similarly suppressed during the high/soft X-ray state in
1996. This is behaviour similar to that shown by GX 339--4 during 1998
(Fender et al. 1999b).

At 15 GHz, a quasi-sinusoidal 3 mJy semi-amplitude modulation is
observed superimposed on the 14 mJy mean level. The minimum occurs at
superior conjunction of the compact body, implying that the
radio-emitting region is associated with this compact object (Pooley
et al. 1999). More recently a further modulation period of 142 days has
been detected (Brocksopp et al. 1999b), explained as precession or
warping of the disk. The source displays a flat spectrum between 2 and
220~GHz, with an average spectral index $|\alpha|\leq 0.15$ (Fender et
al.  2000). A small radio flare coincident with an X-ray flare in 1975
further demonstrated an \mbox{X-ray/radio} correlation (Hjellming \&
Han 1995). Both the observed flat spectrum and the orbital modulation
fit into the slowed adiabatically expanding conical radio jet model
(Hjellming \& Johnston 1988). It is not clear how common the small
flares (less than $\sim$30 mJy) are or exactly how the X-ray state changes
affect the radio emission (Brocksopp et al. 1999b). 

An accretion disk is essential in models of the X-ray
variability. The source has high/soft, low/hard and intermediate
spectral states, spending most time in the low/hard X-ray state. The
X-ray state changes can be modelled using a variable accretion rate;
when the accretion rate is low an advection dominated accretion flow
(ADAF) is postulated (Esin et al. 1998), giving a low radiative
efficiency, as expected for such underluminous quiescent sources.  The
existence of a jet perpendicular to the disk implies that the outflow
process may be important in terms of mass and angular momentum loss from
the disk.

Previous radio maps on both Very Large Array (VLA) (Marti et al.
1996) and MERLIN (Newell 1997) size scales (5 arcsecond and
50~mas beam FWHM respectively) showed no conclusive evidence for
resolved or extended emission.  The VLA image showed emission regions
to the north and south which could be lobes associated with the source
at an extent of $\sim$5 arcminutes.

\section{VLBI and MERLIN observations}

\begin{figure}
\begin{center}
\includegraphics[width=8.5cm]{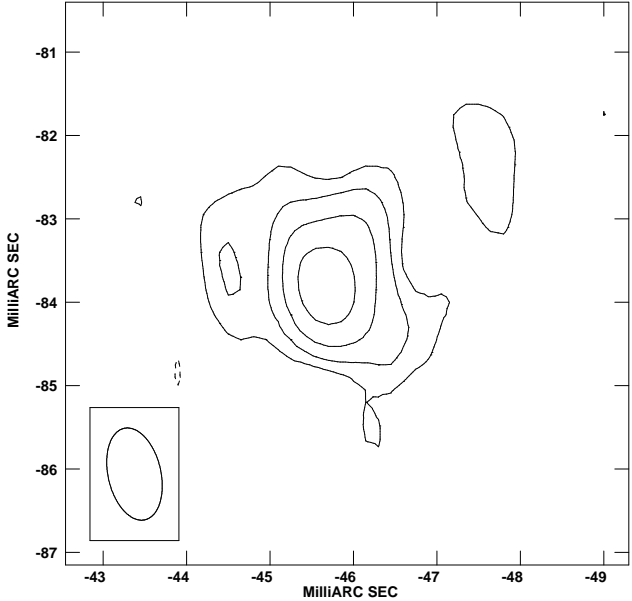}
\end{center}
\caption{\label{cygx1ps} The naturally weighted VLBA image of Cygnus
X-1 from 1997 March 28, at 15.4 GHz has a lowest contour of 0.4 mJy
beam$^{-1}$ and a peak flux density of 5.0 mJy beam$^{-1}$. As with
all images in this paper the contouring represents factors of two.}

\end{figure}

For over a decade Cygnus X-1 has been employed as an astrometric
reference source (Lestrade et al. 1995). Variations in the visibility
amplitude were observed on the very longest transatlantic baselines
(Lestrade, priv. comm.) suggesting resolved source structure on the
milliarcsecond scale. To investigate the milliarcsecond source
structure in more detail, we obtained full track, 15.4 GHz Very Long
Baseline Array (VLBA) observations of the source in 1997.  These were
subsequently followed up in 1998 with multi-epoch and multi-frequency
8.4 and 15.4 GHz observations. The initial 15.4 GHz observations and
the later 8.4 GHz multi-epoch observations are discussed in this
paper. The multi-epoch 15.4 GHz observations will be presented in a
separate paper (de La Force et al. in prep). In this paper we also
present the results of an analysis of archive MERLIN data,
investigating the linear polarization properties of Cygnus X-1.

\subsection{Initial 15.4 GHz VLBA observations}
\label{init}
On 1997 March 28 the VLBA interferometer network was used to observe a
single epoch of Cygnus X-1 at 15.4 GHz while the source was in the
low/hard X-ray state. The run was made at 10:15 -- 19:45 UT (centred
on Julian date 2450901.06), recording at a sustained rate of 128 Mbits
per second. Only one hand of polarization was recorded in Left-hand
Circular Polarization (LCP) providing a total bandwidth of 64 MHz.
The observations were made in phase-reference mode with a typical
cycle time of 2.0 minutes on the phase-calibrator and 2.0 minutes on
the target, Cygnus X-1. The phase-calibrator used throughout these
observations was B1955+335, a bright ($S_{15GHz} \sim$300 mJy),
compact radio source (with an accurately measured astrometric ICRF
position) lying within 1.1 degrees of Cygnus X-1.  The data were
correlated at the NRAO correlator in Socorro, NM, USA.  The
coordinates chosen to correlate the data were those given by the most
recent published radio position, 19$^h$58$^m$21\fs680
+35$^{\circ}$12\arcmin05\farcs887 (epoch 1986, J2000) Lestrade et
al. (1995).  With a total of 8 hours observing time excellent {\it
uv}-coverage of the target source was obtained.
  
The initial analysis of the data was performed with the NRAO {\tt
AIPS} package (e.g. Fomalont 1981). The visibility amplitudes were
calibrated using the system temperatures and gain information provided
by each of the telescopes.  Occasional observations of 3C345 were
used to align the individual 8 MHz sub-bands (IFs) that spanned the
total 64 MHz bandwidth. The antenna residual delays, fringe rates and
gain corrections were determined from the phase calibrator and
interpolated and applied to the target, Cygnus X-1.

Unfortunately, the positions we supplied to the correlator did not
take into account the proper motion of Cygnus X-1. Thus our first map
made from the calibrated Cygnus X-1 data showed no radio emission in
the centre of the map but obvious side-lobe structure from a source
apparently located outside the field was clearly visible. A wide field
image of the region (for which no averaging of the data were
performed) revealed this source to be Cygnus X-1, offset 46 mas to the
east and 84 mas to the south of the nominal phase-centre. Our
position, after correcting for the measured proper motion of the
source (provided by Lestrade, priv. comm.) is within 1.5 mas of the
original astrometric position (Lestrade et al. 1995). The refined
proper motion is 8.7 $\pm$ 0.2 mas yr$^{-1}$ in position angle (PA)
-151 $\pm$ 1$^{\circ}$.

The Cygnus X-1 data were Fourier transformed and then CLEANed using
the {\tt AIPS} task {\sc IMAGR} which allows variable data weighting
using the robust parameter (Briggs 1995). No self-calibration of the
data was performed. A naturally weighted (robust 5) image is presented
in Figure 1. The elliptical Gaussian restoring beam measures $1.13
\times 0.63$ mas in PA 12.9$^{\circ}$.

The source is clearly detected, and is plotted relative to the epoch
1986 coordinates chosen for correlation. The structure is confused and
ambiguous, this is likely due to errors in the phase referencing
process. Unfortunately the source was too weak for reliable
self-calibration with this observing array. An attempt to model fit
these data as a single component with {\tt DIFMAP} (Pearson et
al. 1994) suggested an extension in PA $-15^{\circ} \pm 5$ (Stirling,
Spencer \& Garrett 1998). It seems likely that the radio emission
represents an unresolved radio-jet.

\subsection{Follow up VLBA multi-epoch 8.4~GHz observations}

\subsubsection{Observations and data reduction}

Further high resolution observations of Cygnus X-1 were made in August
1998 with the VLBA and VLA. Three epochs of observations were obtained
every 2 days to cover a large fraction of the period (5.6 days) of the
binary system. The runs were made at 04:19 -- 08:55 UT on the 10th,
(epoch A) at 04:20 -- 08:30 UT on the 12th (epoch B) and at 04:20 --
08:30 UT on the 14th (epoch C). The mid point of each observation was
on Julian date 2451035.76 (A), 2451037.76 (B) and 2451039.76 (C).
Observations were made at 15.4 and 8.4 GHz, the frequencies being
alternated each hour. Only the 8.4 GHz data are considered in this
paper.

Since the source is relatively weak, the phased VLA was also included
as one element in the array. A data rate of 256 Mbits per second was
used (recording LCP in $8\times 8$~MHz sub-bands, 2-bit sampled at the
Nyquist rate). The Cygnus X-1 data were correlated using a position of
19$^h$58$^m$21\fs67634 +35$^{\circ}$12\arcmin05\farcs7940, J2000
(derived from our earlier 15.4 GHz observations but with the source
proper motion now included). The data analysis followed the standard
procedure described earlier in Section 2.1.

\subsubsection{Images and component fluxes}
\begin{figure*}
\begin{center}
\includegraphics[width=8.5cm]{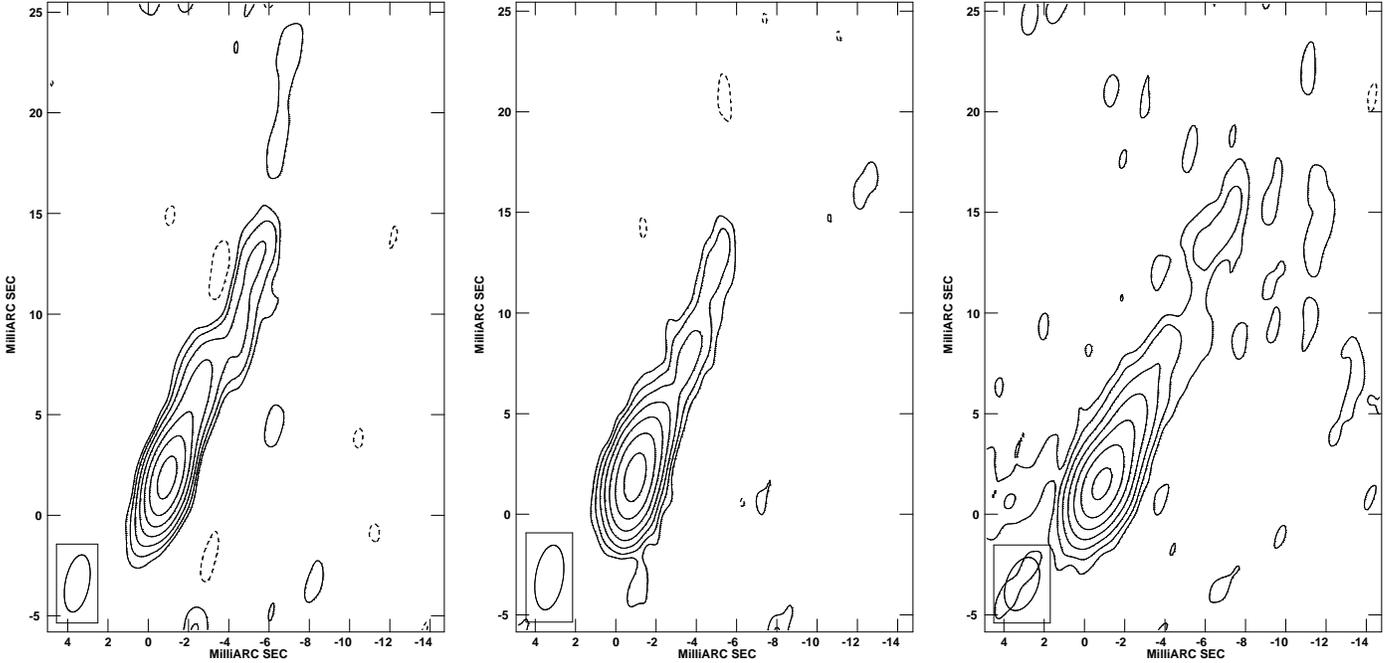}
\end{center}
\caption{\label{cygx1_8} VLBA and phased VLA images of Cygnus X-1 from
August, 1998, at 8 GHz; lowest contour 0.1~mJy. Epoch A is on the left
with peak flux density 8.7 mJy beam$^{-1}$ and convolved with a
Gaussian beam $2.88 \times 1.19$ mas in PA --10.9$^{\circ}$, epoch B
in the middle with peak flux density 9.0 mJy beam$^{-1}$ and convolved
with a Gaussian beam $3.24 \times 1.37$ mas in PA --8.9$^{\circ}$ and
epoch C on the right with peak flux density 7.7 mJy beam$^{-1}$ and
convolved with a Gaussian beam $2.76 \times 1.56$ mas in PA
--20.5$^{\circ}$.}
\end{figure*}

Due to thunder storms and bad weather at the VLA the third epoch had
generally noisier data on the most sensitive VLA--VLBA baselines. The
three naturally weighted images at 8.4 GHz have been self-calibrated in
{\tt AIPS} (using the extremely sensitive `phased' VLA as a reference
antenna) and are presented in Figure~\ref{cygx1_8} with identical
contouring. Here the radio emission from a jet to the north-west is
much clearer, extending out to $\sim$15 mas from the core. The
unresolved core likely represents the base of the visible jet and its
flux density variations with orbital phase are given in
Table~\ref{tab2}.

\begin{table*}
\begin{center}
\begin{tabular}{|c|c|c|c|c|c|c|}
\hline
Epoch & Orbital & Total flux density & Unresolved core & Jet & Off source r.m.s.\\
(1998)& phase&(mJy)      &  (mJy beam$^{-1}$)      & (mJy) &  (mJy beam$^{-1}$) \\
A     & 0.96 &13.0      & 6.9              & 6.1  & 0.04 \\
B     & 0.32 &11.1      & 8.0              & 3.1  & 0.04 \\
C     & 0.68 &11.9      & 6.0              & 5.9  & 0.05 \\
\hline
\end{tabular}
\caption{\label{tab2}Image properties for Cygnus X-1 at 8.4 GHz in
August, 1998. Flux calibration is typically accurate to 5 per cent for
VLBA images at this frequency. The unresolved flux densities per beam
were derived from the uniformly weighted (robust -5) images, which are
not presented in this paper.}
\end{center}
\end{table*}

There is some correlation between orbital ephemeris (from Brocksopp et
al. 1999a) and the flux densities measured from our maps, as the
unresolved core flux density has a maximum around inferior conjunction
(see Table~\ref{tab2}) and the orbital phase modulation of the radio
flux does indeed imply the radio flux variation should originate in or
near the core, as observed. The flare preceding our observations by
around 15 days as shown in Brocksopp et al. (1999b) may complicate the
integrated flux density and period relationship found by Pooley et
al. (1999).

\subsection{MERLIN archival data}
As all of the VLBA observations were made using a single hand of
polarization, it was impossible to examine the linear polarization
emitted from the jet. As MERLIN observes in all four polarization
correlations as standard, linearly polarized emission can be
imaged. Only the total intensity (Stokes I) image was presented by
Newell (1997).

These MERLIN data, observed on 1995 July 19 were retrieved from the
archives and calibrated in full polarization using the MERLIN pipeline
(Thomasson et al. 1993). The 5 GHz observations were made in the
standard MERLIN continuum setup over a bandwidth of 16 MHz. Also
observed were a phase reference source, 1951+355; a flux and
polarization PA calibrator, 3C286; and a point source calibrator,
OQ208. The total intensity map of Cygnus~X-1 was convolved with a
Gaussian beam of $58 \times 37$ mas in PA --81.2$^{\circ}$ and found
to be a point source with a peak flux density of 11.9 $\pm$ 0.6 mJy
beam$^{-1}$ in agreement with Newell (1997). No linear polarization
was detected associated with the source to an upper limit of 0.9 mJy
beam$^{-1}$, ruling out a simple, very ordered field structure.

\section{Analysis of these August 1998 data} 

\begin{figure}
\begin{center}
\includegraphics[width=8.5cm]{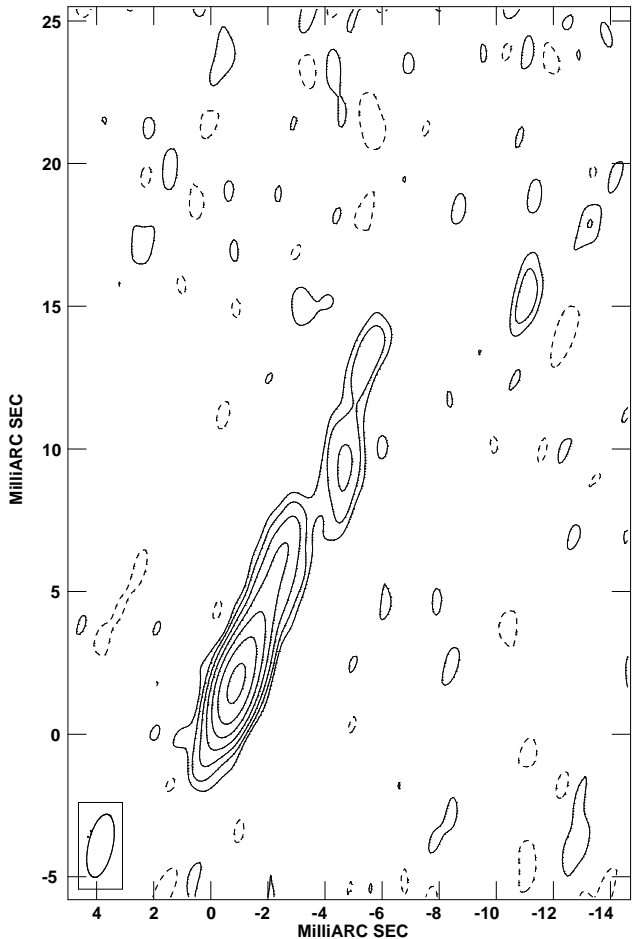}
\end{center}
\caption{\label{hires}A high resolution (robust 0) image of Cygnus
X-1 at 8.4 GHz; lowest contour 0.15~mJy, convolved with a Gaussian beam
$2.25 \times 0.86$ mas in PA --12.4$^{\circ}$.}
\end{figure}

\subsection{Emission mechanism}
We can place a lower limit on the brightness temperature of the
unresolved radio emission simply by using the Rayleigh-Jeans
approximation. Using a flux density of 6 mJy and a solid angle
equivalent to the uniform beam (epoch B) we derive a minimum
temperature of $10^7$~K at 8.4 GHz, inferring that the radiation is
non-thermal, and is not for example due to free-free emission from gas
at 10$^4$~K. The observed spectral index and the detection of
polarization in a number of jet sources suggests that the radiation is
via the synchrotron mechanism. A spectral break is not observed in the
radio or mm regimes, implying that the emitting electrons do not
suffer significant radiative losses (Fender et al. 2000). The flat
spectral index supports the non-thermal interpretation, and is often
attributed to multiple expanding synchrotron-emitting regions within
the beam when observed in the jets from quasars.

\subsection{Doppler boosting}

Following the method of Mirabel \& Rodriguez (1994), under the
assumption of an intrinsically symmetric ejection, then for bulk
motions at a jet velocity $\beta$ the observed ratio of the
approaching jet components flux density, $S_{\rm app}$, to the
receding jet components flux density, $S_{\rm rec}$, (assuming
optically thin components or non-overlapping optically thick
components) is given by

\begin{equation}\label{sides}
\frac{S_{\rm app}}{S_{\rm rec}} = \left(\frac{1+\beta \cos i}{1-\beta
\cos i}\right)^{k-\alpha},
\end{equation}

\noindent with $S_{\nu} \propto \nu^{+\alpha}$; $k=2$ for a continuous
jet and $k=3$ for discrete plasmons; $i$ is the angle to the
line-of-sight. If the source is optically thick in places then the
full radiation transfer equation must be solved, and then
Equation~\ref{sides} gives only an approximation to $\beta$ (Lind \& Blandford 1985), see e.g. Cawthorne (1991) for a derivation of
the (rather small) correction factors for a partially optically thick
and continuous jet.

As we cannot measure the motion of individual components in both the
receding and approaching jets we cannot directly derive both $\beta$
and $i$ and hence the index $k-\alpha$. However we note that the
average accretion disk axis, $i$, (and hence the jet) is likely to be
around 40$^{\circ}$ to the line-of-sight (Bruevich et al. 1978)
although this is not well defined (Brocksopp et al. 1999b). We use the
angle to the line-of-sight to derive $\beta$ for various values of
$k-\alpha$.

We assume that we see only one side of the jet in our epoch A and use
a ratio of approaching and receding flux densities of 50 to derive
lower limits to the jet speed. The ratio is derived by assuming the
receding jet flux is $\leq$ 6 beams $\times$ r.m.s noise per beam,
with 6 beams the extent of the approaching jet. Although the
unresolved flux monitoring suggests a spectral index of around 0,
using --0.6 gives a lower limit to $\beta$ (it is unlikely that a
synchrotron spectrum is much steeper than --0.6). We derive $\beta$
for spectral indices of 0 and --0.6 and for both the continuous and
discrete cases (Table~\ref{calc}). Errors from the poorly constrained
angle to the line-of-sight dominate these calculations and the results
should be regarded as order-of-magnitude only, especially as any
possible attenuation of radiation from the receding jet by external
free-free absorption cannot be quantified with a single frequency. The
derived lower limits to $\beta$ suggest the jet is at least moderately
relativistic $>0.6$ c.

The longer timescale precessional period (142 days) of Cygnus X-1,
where the deviation of the plane of the disk from the orbital plane,
$\delta$, may reach as much as $\sim$37$^{\circ}$ (Brocksopp et
al. 1999b) is a further uncertainty. Soft X-ray monitoring (Brocksopp
et al. 1999b) suggests that the soft X-ray intensity is mid-way
between extrema during our observations; therefore the disk is close
to its average inclination.

\begin{table}
\begin{center}
\begin{tabular}{|c|c|c|}
\hline
Spectral index & Jet type & lower limit to $\beta$ \\
0              & continuous & 0.97 \\
0              & discrete   & 0.74 \\
$-0.6$           & continuous & 0.83 \\
$-0.6$           & discrete   & 0.63 \\
\hline
\end{tabular}
\caption{\label{calc}Lower limits to the jet velocity as a fraction of
$c$ for various jet compositions.}
\end{center}
\end{table}

 The emitted synchrotron intensity of the jet scales with the Doppler
factor, $D$ (e.g. Cawthorne 1991), to a power $\geq$ $(2-\alpha)$. For
$\beta$ of 0.75 and $i=40^{\circ}$ we find $D=1.55$. Using the
observed long term variations in total flux of $\leq$ 3 mJy gives a
maximum value for $\delta$ (assuming, say, $\alpha=0$ for the radio
jet) of only 4$^{\circ}$. Therefore either the jet velocity is
non-relativistic (and hence $D \sim 1$) or the actual precession of
the jet is small compared to that suggested for the soft X-ray
emitting region of the disk. Given our lower limits to $\beta$ (in
Table~\ref{calc}) we suggest that the inner part of the disk may be
precessing at significantly less than the angle of around
$\sim$37$^{\circ}$ suggested by soft X-ray monitoring. One solution to
this would be a twisted accretion disk as postulated for SS433 (Sharp
et al. 1984). Alternatively the $\sim$142 day `period' is a timescale
for changes in the accretion rate, as is likely in GX 339--4 (Corbel
et al. 2000). The extended emission on VLA scales (Marti et al. 1996)
may be consistent with our measured jet position angle, particularly
if the several mas per year south-west proper motion is considered.

\subsection{A continuous and bending jet?}

The higher resolution (robust 0 or partial uniform weighting) image in
Figure~\ref{hires} was made using epoch A, the best set of 8.4 GHz
data. Even with a resolving beam significantly smaller than that
provided by natural weighting the jet still appears continuous in
Figure~\ref{hires}, although with small variations along it and a
general decrease in brightness along the jet. Even at this resolution
we may find that we are not resolving multiple discrete components (as
was not discovered in SS433 until detailed VLBI observations were
made).

Epoch A shows a pronounced bend or kink at around 7 mas from the radio
core. The jet region at 8--15 mas subtends a similar position angle
from the core (21--24$^{\circ}$) in all 3 epochs, whereas the inner
jet ($<$ 7 mas) is at 17$^{\circ}$ during epoch A. The apparent
change in direction of the jet suggests that the position angle of
ejection is varying, perhaps either jet precession or jitter, both of
which are observed in SS433 (Margon \& Anderson 1989). It is also
possible that the kinks observed in the extended emission are
consistent with a bending jet flow, rather than a ballistic motion,
although apparently unrelated to the direction of the proper motion of
Cygnus X-1 as would be expected from ram pressure
considerations. Projection effects will enhance the appearance of a
jet bend (as with apparent 90$^{\circ}$ bends in blazar
jets). The higher resolution image in Figure~\ref{hires} shows that the
bend occurs in a minimum of surface brightness. This is consistent
with an underlying helical magnetic field in which the magnetic field
pitch angle reverses sign at the bend in radio structure e.g. Laing
(1981); Papageorgiou (priv. comm.). Measurements of any linear
polarization could confirm this model.

The bend is no longer apparent in epochs B and C, suggesting that the
feature disappears on a timescale of $\le$ 2 days. This could be due
to rapid motion and decay via adiabatic expansion, or by synchrotron
losses. In the latter case the magnetic field in the component would
need to be greater than 16 G, rather higher than is commonly deduced
for microquasar jets.

It is also apparent in Figure~\ref{hires} that the jet is not resolved
perpendicular to the flow. Using a beam size of 1 mas across the jet
and a maximum distance from the core of 15 mas on the plane of the
sky, the opening semi-angle of the radio-emitting material is
constrained to $\leq$2$^{\circ}$.

The flat radio spectrum in Cygnus X-1 specifically has been modelled
as slowed expansion in a continuous jet (Hjellming \& Johnston
1988). The predicted spectrum for this geometry is flat until a
turnover at around 10~GHz for an inclination of 40$^{\circ}$. However
the self-absorption in a slowed expanding jet model cannot explain the
observed flat spectrum to mm wavelengths (Fender et al. 2000). In
comparison jet components from GRS 1915+105 are observed to become
optically thin less than a few minutes after ejection at mm
wavelengths (Mirabel et al. 1998). Possibly another inverted spectrum
component (such as the disk or stellar wind) as well as the
partially-self-absorbed conical jet is required to explain the
wavelength dependence up to mm wavelengths of the integrated spectrum.

\section{Conclusions}

Cygnus X-1 has been detected with the VLBA at 15.4 GHz, giving a 13
year timebase for proper motion measurements. Subsequent imaging at
8.4 GHz shows that a relativistic jet with either a bending flow or a
variable ejection angle exists in Cygnus X-1. The jet appears to be
narrow ($<2^{\circ}$ opening angle). The source was known to be in a
low/hard X-ray state at the times of all our observations.  Given the
documented X-ray/radio correlations, and our three separate images of
extended radio emission, it is likely that a jet always exists in the
`quiescent' low/hard X-ray state thought to harbour a central ADAF. It
is unclear how the low-level radio flaring of the low/hard X-ray state
fits into the picture of a steady-state accretion disk input/output
system.

Further VLBI observations with a global VLBI array were undertaken in May
2001. A sequence of images within a day should show if the radio jet is
continuous or consists of moving discrete components.

\subsection*{Acknowledgments}

We wish to thank Guy Pooley for flux monitoring efforts and J-F
Lestrade for providing accurate proper motion data. Discussion with
Tim Cawthorne, Denise Gabuzda and Zsolt Paragi improved the scope of
this paper significantly.  MERLIN is a national facility operated by
the University of Manchester on behalf of PPARC. We thank Anita
Richards for retrieval and initial calibration of the MERLIN archive
data. The NRAO is operated by Associated Universities Inc on behalf of
the National Science Foundation. CJD and AMS ackowledge support from
PPARC studentships during the bulk of this work. AMS is currently
supported by a PPARC research grant.

\bsp

\end{document}